\documentclass[%
 reprint,
 amsmath,amssymb,
prl,
 longbibliography,
 lengthcheck,%
]{revtex4-1}

\usepackage{graphicx}
\usepackage{dcolumn}
\usepackage{bm}
\usepackage{hyperref}
\usepackage{txfonts}

\makeatletter

\newcommand{\Rmnum}[1]{\expandafter\@slowromancap\romannumeral #1@}
\makeatother

\begin{document}

\preprint{APS/123-QED}

\title{Microstructure and magnetic anisotropy of electrospun Cu$_{1-x}$Zn$_x$Fe$_2$O$_4$ nanofibers: \\ A local probe study}

\author{Zhiwei Li}\email{z.w.lee@live.cn}
\author{Weiwei Pan}
\author{Junli Zhang}
\author{Haibo Yi}
 \affiliation{Institute of Applied Magnetics, Key Lab for Magnetism and Magnetic Materials of the Ministry of Education, Lanzhou University, Lanzhou 730000, Gansu, P.R. China.}

\date{\today}

\begin{abstract}
Understanding the phenomena at the nanometer scale is of fundamental importance for future improvements of desired properties of nanomaterials. We report a detailed investigation of the microstructure and the resulting magnetic anisotropy by magnetic, transmission electron microscope (TEM) and M\"ossbauer measurements of the electrospun Cu$_{1-x}$Zn$_x$Fe$_2$O$_4$ nanofibers. Our results show that the electrospun Cu$_{1-x}$Zn$_x$Fe$_2$O$_4$ nanofibers exhibit nearly isotropic magnetic anisotropy. TEM measurements indicate that the nanofibers are composed of loosely connected and randomly aligned nanograins. As revealed by the Henkel plot, these nanofibers and the nanograins within the nanofibers are dipolar coupled, which reduces the effective shape anisotropy leading to a nearly random configuration of the magnetic moments inside the nanofibers, hence, the observed nearly isotropic magnetic anisotropy can be easily understood.

\begin{description}
\item[PACS numbers]
81.07.-b, 75.75.+a
\end{description}
\end{abstract}

\maketitle


\section{\label{sec:Intro}Introduction}

Nanostructured materials find themselves potential applications in many fields such as ultrahigh-density data storages, sensors, drug delivery systems, and high-frequency devices \cite{App-I,App-II,App-III,App-apl} due to their distinctive properties that are not realized by their bulk counterparts. Especially, one dimensional nanowires or nanofibers (NWFs) \cite{App-I,OneDBiFeO3,OneDNi,OneDLaPbMnO} have recently received much interest not only because their potential usages but also because their fundamental importance from a theoretical point of view \cite{AnisotropyM,EDequation-prb}. Among various preparation methods, electrospinning has been proved to be an efficient process that can fabricate polymer NWFs on an industrial scale \cite{BioAdv-Review,Polymer-Rev}. And, during the last decade, remarkable progress has been made in applying this technique to the fabrication of magnetic NWFs \cite{Jap-Rev}.

Spinel ferrites NWFs are always an important subject of many research groups. And during the past few years, TMFe$_2$O$_4$ (TM=Ni,Co,Mg,Mn, etc.) ferrite nanofibers have been synthesized using electrospinning and their magnetic properties were also investigated in detail \cite{pan7,pan8,pan9,pan10}. Copper ferrite is also an interesting material that has been widely used in many areas \cite{CuPan12,CuPan11}. In our previeous work, Cu$_{1-x}$Zn$_x$Fe$_2$O$_4$ nanofibers have been prepared using electrospinning technique and the influence of Zn$^{2+}$ substitution on crystal structure, morphology and magnetic properties have been investigated \cite{WWPan}. Usually, if the shape anisotropy dominates over the magnetocrystalline anisotropy, the easy magnetization direction should along the long axis of a NFW \cite{KsKm}. Interestingly, however, we have found that the magnetic easy axis is not along the long axis of the nanofibers in sharp contrast with the usual sense.

In the present study, in order to get a better understanding of the unusual magnetic anisotropy of the Cu$_{1-x}$Zn$_x$Fe$_2$O$_4$ nanofibers, the microstructure of these nanofibers have been studied in detail by TEM and M\"ossbauer spectroscopy. The magnetic interactions of the sample were also investigated using the Henkel plot. And the present results reveal dipolar interactions of our nanofibers, which reduces the effective shape anisotropy and well explains the observed unusual magnetic anisotropy.

\section{\label{sec:Experiment}Experiments}
Cu$_{1-x}$Zn$_x$Fe$_2$O$_4$ (x=0$\sim$1.0) nanofibers were synthesized by electrospinning combined with Sol-Gel technique \cite{Jap-Rev,WWPan}. Phase purity was checked by X-ray diffraction (XRD) measurement using a Philips X'pert diffractometer with Cu $K_{\alpha}$ radiation. Morphology examination and elemental analysis were performed using a transmission electron microscope (TEM, FEI Tecnai F30) and a field emission scanning electron microscope (SEM, Hitachi S-480) equipped with an energy-dispersive X-ray spectrometer (EDXS). DC magnetic properties were characterized using a vibrating sample magnetometer (VSM, Lakeshore 7403, USA). We will describe the details of the sample preparation, structural, and the static magnetic properties elsewhere \cite{WWPan}. Transmission M\"ossbauer spectra (MS) were recorded at room temperature using a conventional constant acceleration spectrometer with a $\gamma$-ray source of 25\,mCi $^{57}$Co in palladium matrix. The isomer shift quoted in this work are relative to that of the $\alpha$-Fe.

\begin{figure}[htp]
\includegraphics[width=8 cm]{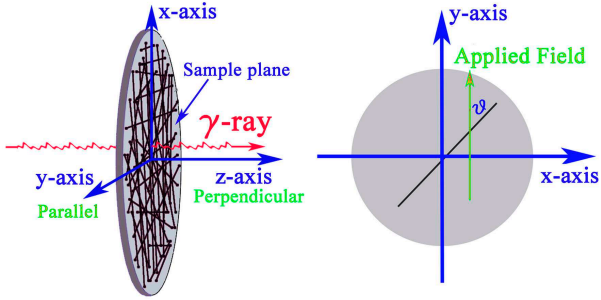}
\caption{\label{Measure}(Color online) Schematic diagram for the DC magnetic and M\"ossbauer measurements (left) and front view of the sample plane (right) (see text).}
\end{figure}

For the DC magnetic and M\"ossbauer characterization, $\sim$30\,mg of the sample were weighted and then uniformly deposited on a thin nonmagnetic underlayer. From the SEM and TEM results we know that the nanofibers have a very large length to diameter ratio. So, the nanofibers should lie parallel to the sample plane during the measurements as illustrated in Fig. \ref{Measure}. In this case, the incident $\gamma$-rays should perpendicular to the long axis of the nanofibers, Fig. \ref{Measure} (a). For DC magnetic measurements, two different hysteresis loops were recorded with the magnetic field, H, applied in the y-axis direction (H should be parallel to the sample plane) and with H applied in the z-axis direction (H should be perpendicular to sample plane), respectively.

\section{\label{sec:Results}Results and Discussion}

\begin{figure}[htp]
\includegraphics[width=8 cm]{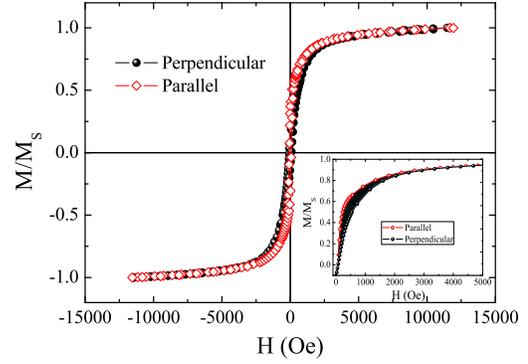}
\caption{\label{VSM}(Color online) Typical hysteresis loops of Cu$_{0.6}$Zn$_{0.4}$Fe$_2$O$_4$ nanofibers. The measurements were done with magnetic field applied parallel (red open diamond) and perpendicular (black solid circle) to the sample plane, respectively. The right-lower inset is an enlargement of the low-field part of the magnetization curve and the shaded area indicates the difference between the two curves.}
\end{figure}

In Fig. \ref{VSM}, we present the hysteresis loops of Cu$_{0.6}$Zn$_{0.4}$Fe$_2$O$_4$ nanofibers. The lower inset is an enlargement of the low-field part of the magnetization curve. As described in the experimental section, parallel and perpendicular indicates that the hysteresis loops were measured with the magnetic field applied parallel and perpendicular to the sample plane, respectively. Usually, if the shape anisotropy dominates over the intrinsic magnetocrystalline anisotropy, an easy magnetization axis along the long axis of a nanowire is expected \cite{KsKm}. For the Cu$_{0.6}$Zn$_{0.4}$Fe$_2$O$_4$ nanofibers, the saturation magnetization is measured to be $M_S$=58.4\,emu\,g$^{-1}$, leads to an expected shape anisotropy of $K_{shape}=3.2\times10^5$\,erg\,cm$^{-3}$, which is much larger than the magnetocrystalline anisotropy for bulk copper ferrites, $\sim$0.6$\times10^5$\,erg\,cm$^{-3}$ \cite{HandbookMV8}. This means that the expected easy axis should be along the nanofibers. Interestingly, one can see that the two hysteresis loops almost overlap. Same behavior of the hysteresis loops have also been found for other compositions, suggesting that the electrospun Cu$_{1-x}$Zn$_{x}$Fe$_2$O$_4$ nanofibers have a smaller shape anisotropy than expected.

To examine the anisotropy more clearly, we now estimate the effective anisotropy field and compare it with experiments quantitatively. As is well known, the experimentally measured shape anisotropy field for nanowires can be expressed as $H_d=-NM_S$, where $N=1/2$ is the demagnetization factor and $M_S$ is the saturation magnetization. In our case, however, the experimentally measured effective anisotropy field, $H_{eff}$, should be different since the nanofibers are randomly aligned within the sample plane. In order to calculate $H_{eff}$, we first deduce the effective demagnetization factor for a disk-like sample composed of randomly aligned nanofibers. We define $\theta$ to be the angle between the applied magnetic field and the long axis of a nanofiber, Fig. \ref{Measure} (b). Then, the demagnetization energy can be expressed as \cite{EDequation-prb}
\begin{eqnarray}
\label{Ed}
E_d=\frac{1}{2}\mu_0N_{\theta}M^2=\frac{1}{2}\mu_0(N_{\parallel}M_{\parallel}^2+N_{\perp}M_{\perp}^2) \\ \nonumber
=\frac{1}{2}\mu_0N_{\perp}(M_S\sin\theta)^2,
\end{eqnarray}
where $\parallel$/$\perp$ indicates parallel/perpendicular to the long axis of the nanofiber. Rewrite equation (\ref{Ed}) as $E_d=\frac{1}{2}\mu_0(N_{\perp}\sin^2\theta)M_S^2$, one could get the demagnetization factor for a nanofiber along the $\theta$ direction, $N_{\theta}=N_{\perp}\sin^2\theta=\frac{1}{2}\sin^2\theta$. For a disk-like sample composed of uniformly aligned nanofibers, the effective demagnetization factor can be determined to be $N_{eff}=1/4$ by integrating $N_{\theta}$ over all possible directions. Then, the effective anisotropy field is expressed as
\begin{eqnarray}
H_{eff}=|-N_{\perp}M_S - (-N_{eff}M_S)|=\frac{1}{4}M_S.
\label{Heff}
\end{eqnarray}
Using the measured $M_S$=58.4\,emu\,g$^{-1}$, we can calculate the effective anisotropy field to be $H_{eff}$=1006\,Oe. This value is much larger than the measured value from the hysteresis loops, $H_k\sim$600\,Oe, which is deduced as twice of the shaded area \cite{Heff-loops} indicated in the lower inset of Fig. \ref{VSM}. This suggests that the effective anisotropy is rather different from that expected for nanowires that are dominated by the shape anisotropy. In other words, if we consider one single nanofiber, the easy magnetization direction is not along the long axis of the fiber, which is evidenced by our M\"ossbauer measurements. The different magnetic anisotropy was also observed by R.C. Pullar and A.K. Bhattacharya \cite{refRCPullar} in randomly oriented M hexa-ferrite fibers, where they show that alignment effects play important roles in the regarding anisotropy effect.

\begin{figure}[htp]
\includegraphics[width=8 cm]{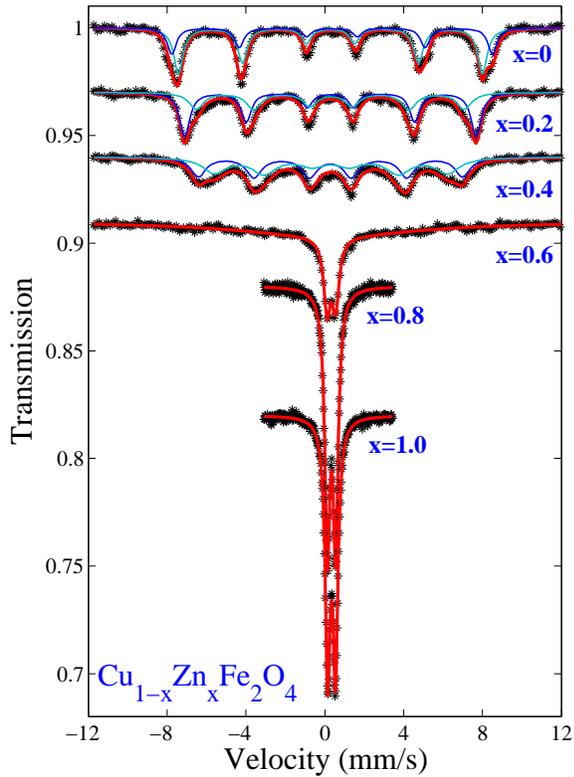}
\caption{\label{Moss}(Color online) M\"ossbauer spectra of the Cu$_{1-x}$Zn$_x$Fe$_2$O$_4$ (x=0$\sim$1.0) nanofibers taken at room temperature. The spectra were fitted with two sextets for x=0, 0.2 and 0.4, with only one doublet for x=0.8 and 1.0.}
\end{figure}

To understand the above observed unusual anisotropy effect, M\"ossbauer measurements have been employed to probe the microscopic spatial distribution of the magnetic moments inside the nanofibers. Fig. \ref{Moss} shows the room temperature $^{57}$Fe M\"ossbauer spectra of the Cu$_{1-x}$Zn$_x$Fe$_2$O$_4$ (x=0$\sim$1.0) nanofibers. The fitted hyperfine parameters are given in Table \ref{TableMP}. Clearly, the spectrum evolves from two sextets (corresponding to the Fe atoms sit at tetrahedral (A) sites and octahedral (B) sites, respectively) for x=0 to a quadrupole split doublet for x=0.6, indicating the collapse of the long range magnetic order upon Zn substitution, coincidence with magnetization measurements \cite{WWPan}. From the shape of the spectra one can see that the sextets broadens, instead of the superposition of the sextets and the doublet, as x increases till the collapse point for x=0.6, which is indication of homogeneous Zn substitution of the Cu ions.

\begin{table}[ht]
\centering
\caption{Hyperfine parameters of Cu$_{1-x}$Zn$_x$Fe$_2$O$_4$ (x=0$\sim$1.0) nanofibers extracted from least squares fit of the M\"ossbauer spectra shown in Fig. \ref{Moss}. $\delta$ denotes the isomer shift, $\Delta E_Q$ is the quadrupole splitting and $B_{hf}$ is the hyperfine magnetic field.}
\label{TableMP} {
\begin{tabular}{l l l l l}
\hline \hline
Sample  &    Site   &  $\delta$ (mm\,s$^{-1}$)         &   $\Delta E_Q$ (mm\,s$^{-1}$) &  $B_{hf}$ (T)     \\
\hline
x=0.0    &   A       &   0.281(10)   &   -0.024(10)  &   47.97(11)         \\
        &   B       &   0.365(17)   &   -0.012(17)  &   50.86(16)       \\
x=0.2   &   A       &   0.288(12)   &   -0.001(12)  &   45.82(12)       \\
        &   B       &   0.305(29)   &   -0.025(29)  &   42.15(45)       \\
x=0.4   &   A       &   0.288(22)   &   -0.012(22)   &   41.44(34)         \\
        &   B       &   0.325(45)   &   -0.032(41)  &   36.29(81)       \\
x=0.6   &   -       &   0.335(10)   &   0.439(14)  &   -      \\
x=0.8   &   -       &   0.342(2)   &   0.416(4)  &   -         \\
x=1.0   &   -      &   0.346(2)   &   0.393(3)  &   -       \\
\hline \hline
\end{tabular}}
\end{table}

To exclude the spectral line broadening effect on the relative intensity ratio of the six lines, we take CuFe$_2$O$_4$ as an example to investigate the microscopic spatial distribution of the magnetic moments inside the nanofibers. In M\"ossbauer measurements, as is well known, the direction of the magnetic moments can be deduced by the relative intensities of the six absorption lines of the Zeeman split sextet. The relative intensities as a function of the polar angle $\theta$ between the magnetic moment and the incident $\gamma$-rays is expressed as \cite{Moss-book}
\begin{eqnarray}
\frac{I_{2,5}}{I_{1,6}}=\frac{4\sin^2\theta}{3(1+\cos^2\theta)},
\label{IntensityR}
\end{eqnarray}
where $I_{2,5}$ and $I_{1,6}$ are the line intensities of the 2,5th and 1,6th peaks of the magnetic splitting sextet, respectively. The best fit of our M\"ossbauer spectra for CuFe$_2$O$_4$ yields $I_{2,5}/I_{1,6}=2.25$, which is close to the value of 2 for samples with randomly aligned magnetic moments. This is in sharp contrast with the usual expected value of 4 for nanowires where the magnetic moments usually aligns parallel to the long axis of the nanowires due to the shape anisotropy. This result explains well why the measured effective anisotropy is much smaller than expected.

\begin{figure}[htp]
\includegraphics[width=8 cm]{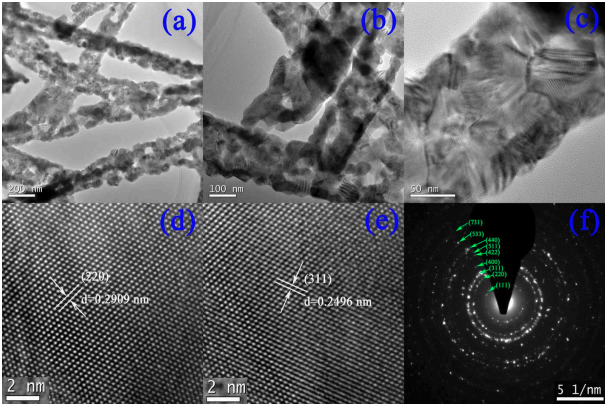}
\caption{\label{TEM}(Color online) TEM micrographs of the CuFe$_2$O$_4$ nanofibers (a)-(c), high magnification TEM images of the nanofibers showing the crystalline lattice structures (d) and (e), and the corresponding SAED pattern of a single fiber (f).}
\end{figure}

TEM has been employed to further examine the microstructures of the CuFe$_2$O$_4$ nanofibers. In Fig. \ref{TEM} (a)-(c), we presents the TEM micrographs of the nanofibers, from which one can see that the nanofibers are about $\sim$200\,nm in diameter. A closer look of the nanofibers reveal that these fibers are composed of smaller grains of about $\sim$50\,nm in size and are loosely connected with each other. Crystalline lattice structures are clearly shown in the high magnification TEM images, Fig. \ref{TEM} (b) and (c), indicating good crystallinity of our sample.  Fig \ref{TEM} (f) shows the selected area electron diffraction (SAED) pattern of a single fiber. All the bright rings can be indexed to the cubic spinel structure. Another important fact one can learn from the bright rings of the SAED pattern is that the small grains are randomly oriented in the fiber.

As we know, the NWFs are usually composed of closely connected smaller grains, $\sim$ several nanometers. And according to Herzer's statement \cite{HerzerStatement}, when the grain size ($D$) along with the intergranular distance ($S$) is smaller than the exchange length ($L_{ex}=\sqrt{2A/(\mu_0M_S^2)}$), the exchange coupling takes place. The magnetocrystalline anisotropy will be suppressed by the exchange coupling, $\langle K\rangle=K_1/\sqrt{N}$, where $K_1$ is the magnetocrystalline anisotropy constant and $N$ is the granular amount within the volume of $V=L_{ex}^3$. This will effectively increase the demagnetization effect, which favors that the magnetization align parallel to the long axis of the nanowires. If we adopt the value of $A=1.2\times10^{-11}$\,J\,m$^{-1}$ \cite{SimulationJMMM}, the estimated exchange length is $L_{ex}\sim$33.8\,nm. In the present nanofibers, as learned from TEM results, the size of the small grains ($\sim$50\,nm) exceeds this characteristic length. Thus, dipolar interaction should be expected in the present Cu$_{1-x}$Zn$_x$Fe$_2$O$_4$ nanofibers, which means that the magnetization in two neighboring grains favor an antiparallel alignment.

To prove our above arguments, we further study the magnetic interactions of the small grains inside the nanofibers. A useful tool for studying the interactions among magnetic nanograins is the Henkel plot or the $\delta m(H)$ plot \cite{Henkel1,Henkel2}. It is well known, for an assembly of magnetically noninteracting nanograins, that the isothermal remanent magnetization $M_r(H)$ and the DC demagnetization remanence $M_d(H)$ should obey the Stoner-Wohlfarth relation:
\begin{eqnarray}
M_d(H) = M_r(\infty) - 2M_r(H),
\label{SWRelation}
\end{eqnarray}
where $M_r(\infty)$ is the maximum remanent magnetization measured after saturation. Henkel first proposed that the deviation of the measured $M_d(H)$ and $M_r(\infty)-2M_r(H)$ can be used to study the magnetic interactions in real systems. The Henkel plot is expressed as follows \cite{Henkel1}:
\begin{eqnarray}
\delta m(H) = \frac{M_d(H) - (M_r(\infty) - 2M_r(H))}{M_r(\infty)}.
\label{SWRelation}
\end{eqnarray}
Positive values of $\delta m(H)$ are due to exchange interactions promoting the magnetized state, while negative values of $\delta m(H)$ correspond to dipolar interactions tending to assist magnetization reversal.

\begin{figure}[htp]
\includegraphics[width=8 cm]{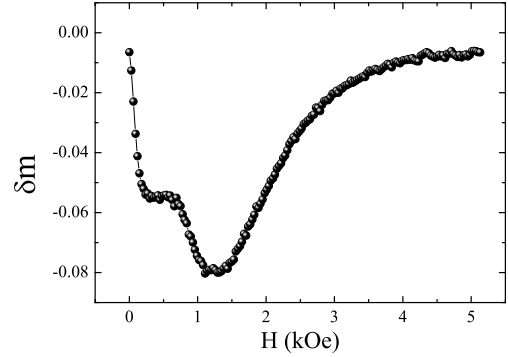}
\caption{\label{deltam} $\delta m(H)$ plot of the CuFe$_2$O$_4$ nanofibers.}
\end{figure}

Typical room temperature $\delta m(H)$ plot for CuFe$_2$O$_4$ nanofibers is shown in Fig. \ref{deltam}. As can be seen, two minimum values of the $\delta m(H)$ curve that possibly due to inter-fiber and inter-grain interactions are observed \cite{DipolarI-1,DipolarI-2,DipolarI-3}. Negative values of $\delta m(H)$ are indicative of a dipolar type of interaction, which is expected in the case of loosely connected nanograins. Thus, we can conclude that the present nanofibers are composed of dipolar coupled nanograins and the interaction between these nanofibers is also dipolar type, which tend to assist the magnetization reversal process hence reduce the effective shape anisotropy and lead to the above observed unusual magnetic anisotropy effect.

\section{\label{sec:Conclusion}Concluding remarks}
In summary, we have studied the microstructure and magnetic anisotropy effect of electrospun Cu$_{1-x}$Zn$_x$Fe$_2$O$_4$ nanofibers. Magnetic and M\"ossbauer measurements reveal that the electrospun nanofibers have a rather isotropic magnetic anisotropy. This is in sharp contrast with the usual sense for nanowires that prepared by conventional methods. From the negative values of the $\delta m(H)$ plot as well as the TEM results, we can conclude that the present nanofibers are composed of randomly aligned and dipolar coupled nanograins and these nanofibers are also dipolar coupled, which reduces the effective shape anisotropy and well explains the observed unusual magnetic anisotropy effect.

\begin{acknowledgments}
The author (Zhiwei Li) is grateful to Dr. X. Yang for useful discussions. This work was supported by the National Natural Science Foundation of China under Grants No. 10774061.

\end{acknowledgments}


\end{document}